\documentclass[5p,authoryear]{elsarticle}
\usepackage{natbib}	
\usepackage{lineno}
\usepackage[colorlinks]{hyperref}
\usepackage[dvipsnames]{xcolor}
\journal{Journal of \LaTeX\ Templates}

\begin{document}

\begin{frontmatter}

\title{Visible Analysis of NASA Lucy Mission Targets Eurybates, Polymele, Orus and Donaldjohanson.\tnoteref{mytitlenote}}
\tnotetext[mytitlenote]{Based on observations obtained at the Southern Astrophysical Research (SOAR) telescope, which is a joint project of the Minist\'erio da Ci\^encia, Tecnologia, Inova\c{c}\~ao e Comunica\c{c}\~oes (MCTIC) da Rep\'ublica Federativa do Brasil, the U.S. National Optical Astronomy Observatory (NOAO), the University of North Carolina at Chapel Hill (UNC), and Michigan State University (MSU), in observations at GTC (ORM-La Palma) and, in observations at the Observat\'orio Astron\^omico do Sert\~ao de Itaparica (OASI, Itacuruba) of the Observat\'orio Nacional (ON/MCTIC, Brazil).}

\author[mymainaddress,mysecondaryaddress]{A.C. Souza-Feliciano\corref{mycorrespondingauthor}}
%\ead{carolofisica@gmail.com}
\author[mysecondaryaddress]{M. De Pr\'a}
\author[mysecondaryaddress]{N. Pinilla-Alonso}
\author[mymainaddress]{A.Alvarez-Candal}
\author[mysecondaryaddress]{E. Fern\'andez-Valenzuela}
\author[mythreeadress,myfouradress]{J. de Le\'on}
\author[myfiveadress]{R. Binzel.}
\author[mymainaddress]{P. Arcoverde}
\author[mymainaddress]{E. Rond\'on}
\author[mymainaddress]{M. E. Santana}

\address[mymainaddress]{Observat\'orio Nacional, Rio de Janeiro, 20921-400, Brazil.}
\address[mysecondaryaddress]{Florida Space Institute, University of Central Florida, Florida, USA.}
\address[mythreeadress]{Instituto de Astrof\'isica de Canarias. C/V\'ia L\'actea s/n, E-38205 La Laguna, Tenerife, Spain.}
\address[myfouradress]{Departamento de Astrof\'isica, Universidad de La Laguna. E-38206 La Laguna, Tenerife, Spain.}
\address[myfiveadress]{Department of Earth, Atmospheric, and Planetary Sciences, Massachusetts Institute of Technology, Cambridge, MA 02139, USA.}

\begin{abstract}
%This template helps you to create a properly formatted \LaTeX\ manuscript.
Jupiter Trojan asteroids are minor bodies that share Jupiter's orbit around the Sun. Although not yet well understood in origin and composition, they have surface properties that, besides being comparable with other populations of small bodies in the Solar System, hold information that may restrict models of planetary formation. Due their importance, there has been a significant increase in an interest in studying this population. In this context arises the NASA Lucy Mission, with a planned launch of 2021. The Lucy Mission will be the first one to address a group of 6 objects (five Jupiter Trojan and one main belt asteroid) with the aim of investigating, in detail, their nature. In order to provide valuable information for mission planning and maximize the scientific return, we carried out ground based observations of four targets of the mission. Aimed at looking for variabilities on the spectra of (3548) Eurybates, (15094) Polymele and (21900) Orus, we performed rotationally resolved visible spectroscopy of them at SOAR Telescope. We also analyzed the first visible spectrum obtained for the main belt asteroid (52246) Donaldjohanson at Gran Telescopio Canarias. The spectra of (21900) Orus and (15094) Polymele present rather homogeneous characteristics along the surfaces, and their taxa correspond with those of the two dominant populations in the Trojan population, the P- and the D-type group of objects. Spectroscopy of (3548) Eurybates, on the other side, suggests that some variation on the characteristics of the reflectance of this body could be related with its collisional history. Donaldjohanson, the only main belt object in the group of targets, shows, according to our visible spectrum, hints of the presence of hydrated materials. Lucy mission will investigate the surface composition of these targets and will shed light on their connections with other minor bodies populations and in their role on the evolution of the Solar System.
\end{abstract}


\end{frontmatter}

%\linenumbers

\section{Introduction}

Jupiter Trojan asteroids (hereafter, JTs) are minor bodies that share Jupiter's orbit around the Sun in stable orbits over the age of the Solar System \citep{levison1997dynamical, marzari2003stability} in two Lagrange regions, L4 and L5. Their origin and composition are a subject of current debate. \cite{marzari1998capture} proposed that they formed in a region close to Jupiter and hold information of the central part of the solar nebula, while \cite{morbidelli2005chaotic} argued that they were captured from a more distant region, during a phase of the proposed planetary migration period in the early Solar System. Therefore, according to these authors, JTs should contain information from the outermost part of the protoplanetary disk. In this context, the study of the JTs dynamical and physical properties can provide important clues about the history and evolution of our planetary system. %In both cases, the dynamical configuration kept the JTs isolated from the asteroid Main Belt throughout the Solar System history and, in spite case of eventual interactions with other populations of minor bodies as the Hildas, Jupiter Family Comets, and the Centaurs, the collisional evolution of the JTs has been dominated mostly by intrapopulation collisions \citep{marzari97, marzari1998growth}. Therefore, the JTs may be considered primitive asteroids, whose dynamical and physical properties were preserved and can provide importants clues about the history and evolution of our planetary system. 

\begin{table*}[t!]
	\centering
	\small
	\begin{tabular}{lccccc}
		\hline
		\hline 
		Object & Spec. ID & V (mag)& Date UT$_{start}$ & T$_{exp}$(s) &  Airmass \\ 
		\hline
		& I & 17.07 & 08 July 2018 03:00 & 3 x 360 & 1.55 \\ 
		Eurybates & II & 17.07 & 08 July 2018 06:29 & 3 x 360 & 1.01  \\ 
		& III & 17.07 & 08 July 2018 08:32 & 3 x 360 & 1.08 \\
		\hline
		& I &18.82 & 08 July 2018 02:00 & 3 x 600 & 2.00 \\
		Polymele & II & 18.82 &08 July 2018 04:36 & 2 x 600 & 1.10 \\ 
		& III & 18.69 & 12 August 2018 23:56 & 3 x 600 & 1.69 \\
		\hline 
		& I & 16.96 &08 July 2018 01:10 & 3 x 360 & 1.36 \\ 
		& II & 16.95 &08 July 2018 03:40 & 3 x 360 & 1.09  \\  
		Orus & III & 16.95 &08 July 2018 05:56 & 3 x 360 & 1.06\\
		& IV & 16.95 &08 July 2018 07:40 & 3 x 360 & 1.30 \\  
		\hline 
		Donaldjohanson &  & 19.17 & 26 May 2018 21:45 & 3 x 600 & 1.34\\
		\hline 
	\end{tabular}
	\caption{Observational circumstances of the JTs and MBA Lucy's targets. Abbreviations are defined as follows: Spectral identification (Spec. ID), Visual magnitude (V (mag)), Date/Universal time of the observation (Date/UT$_{start}$), and Exposure Time (T$_{exp}$).}
	\label{tab1} 
\end{table*}

Several spectroscopic observations at the visible and near-infrared (NIR) wavelengths revealed featureless spectra \citep{dotto2006surface,fornasier2007visible, emery2010} with colors that vary from gray to red, without signatures of ices or hydrated silicates \citep{de2010peculiar, yang2007, yang2011}. Most of the JTs could be classified in the Tholen taxonomy \citep{tholen1984} as D- or P-types, with a minor fraction of C-types. Studies on thermal wavelenghts showed that they are predominantly low-albedo objects \citep{grav11} similar to primitive asteroids and cometary nuclei. These similarities suggest that the JTs also have a primitive nature, having undergone little to no thermal alteration since their formation.

Further investigations revealed a bimodality in their color distribution both in visible and NIR wavelengths, with one "red" peak related to the D-types, and one less red peak, similar to P-types \citep{emery2010, wong2014, de2018primass}. This bimodality has been interpreted as the predominance of two intrinsically different compositional groups. This hypothesis is strengthened by the work of \cite{Brown2016}, which searched for a feature located at around 3 $\mu m$, diagnostic of water ice, and hydrated minerals, in the surfaces of JTs. The feature was identified in the spectrum of the P-types JTs, while all D-types presented featureless spectra.  

Due to the importance of the JTs in constraining dynamical models, there has been a significant increase in an interest in studying this population. In this context arises the NASA Lucy Mission, with a planned launch of 2021. The Lucy Mission will be the first one to address a group of objects with the aim of investigating, in detail, their nature. The mission will perform six flybys: one on the Main Belt Asteroid (MBA) (52246) Donaldjohanson, and five on JTs (3548) Eurybates, (15094) Polymele, (11351) Leucus, (21900) Orus and the binary (617) Patroclus-Menoetius (hereafter refereed only by the objects' name). The mission targets were selected to cover a range of common taxonomic types among the JTs with distinct physical properties. Polymele is a small P-type asteroid with approximatelly 21 km in diameter, while the binary system Patroclus-Menoetius have more than 100 km each in diameter and are also P-types objects. On the other hand, Orus and Leucus are D-types objects, with intermediate diameters (51 and 34 km, respectively) and rotation period completely differents: Orus has a rotation period of $\sim$ 14 hours and Leucus almost 450 hours \citep{buie2018light}. Eurybates and Donaldjohanson are both members of primitive dynamical families, the first in the L4 swarm and the other in the main belt. Eurybates is the largest member of the homonyms family with a diameter of $\sim$ 63 km, and Donaldjohanson is a small body, $\sim$ 4 km, of the Erigone Family \citep{nesvorny2015}. These objects were selected as representatives of the JTs population. The investigation of their properties will reveal the diversity among the group and will help to interpret the whole population providing constraints to the formation and evolution of the Solar System.

\begin{table}[t!]
	\centering
	\small
	\begin{tabular}{cccc}
		\hline
		\hline
		Star & Date/UT$_{start}$ & T$_{exp}$(s) & Airmass\\
		\hline
		 & 07 July 2018 23:25 & 3 x 30 & 1.31\\
		SA 107-684 & 08 July 2018 00:28 & 3 x 30 & 1.19\\
		 & 13 August 2018 01:30 & 3 x 120 & 1.48 \\
		 \hline
		 & 26 May 2018 22:16 & 3 x 2 & 1.34\\
		SA 107-998 & 08 July 2018 00:06 & 3 x 40 & 1.22\\
		 & 12 August 2018 23:03 & 3 x 90 & 1.16 \\
		 & 13 August 2018 01:17 & 3 x 90 & 1.43 \\
		 \hline
		 & 08 July 2018 08:09 & 3 x 75 & 2.16\\
	SA 110-361 & 12 August 2018 23:34 & 3 x 120 & 1.45 \\
		 & 13 August 2018 01:44 & 3 x 120 & 1.17 \\
		 \hline
		SAO & 08 July 2018 10:06 & 2 x 5 & 1.05\\
		\hline
		SA 102-1081 & 26 May 2018 21:35 & 3 x 1 & 1.24\\
		\hline
		SA 112-1333 & 13 August 2018 01:03 & 3 x 90 & 1.64 \\ 
		\hline
	\end{tabular}
	\caption{Observational conditions of the solar analogs stars. Abbreviations are defined as follows: Date/Universal time of the observation (Date/UT$_{start}$) and Exposure Time (T$_{exp}$).}
	\label{tab2}
\end{table}

%falar primeiro da missão, depois de todos os alvos da missão, nossa amostra e o que nós iremos analisar, ultimo da estrutura do paper OK
Aiming to provide valuable information for mission planning and maximizing the scientific return, we carried out an observational study of four of the mission's targets: Donaldjohanson, Eurybates, Polymele and Orus. In this work, we conducted a search for surface heterogeneity in three out of five JTs Lucy targets: Eurybates, Polymele, and Orus. To ensure the robustness of our analysis, we applied two techniques: rotationally resolved spectroscopy, using the Goodman spectrograph at Southern Astrophysical Research (SOAR) Telescope, and to confirm a case of variability, we analyzed the rotational Light-Curve (LC), observed with two photometric filters using the 1-m Telescope of the Observat\'orio Astron\^omico do Sert\~ao de Itaparica (OASI). We characterized, for the first time, the spectrum of the Lucy MBA target Donaldjohanson, obtained at 10.4m Gran Telescopio Canarias (GTC). The paper is organized as follows. In Sect. \ref{dois} we describe the observations and data reduction process, in Sect. \ref{tres} we show the analysis and results of our data, in Sect. \ref{cinco} we discuss the results and in Sect. \ref{seis} we present our conclusions. 

\begin{figure*}[!t]
	\begin{center}
		\includegraphics[scale=0.40]{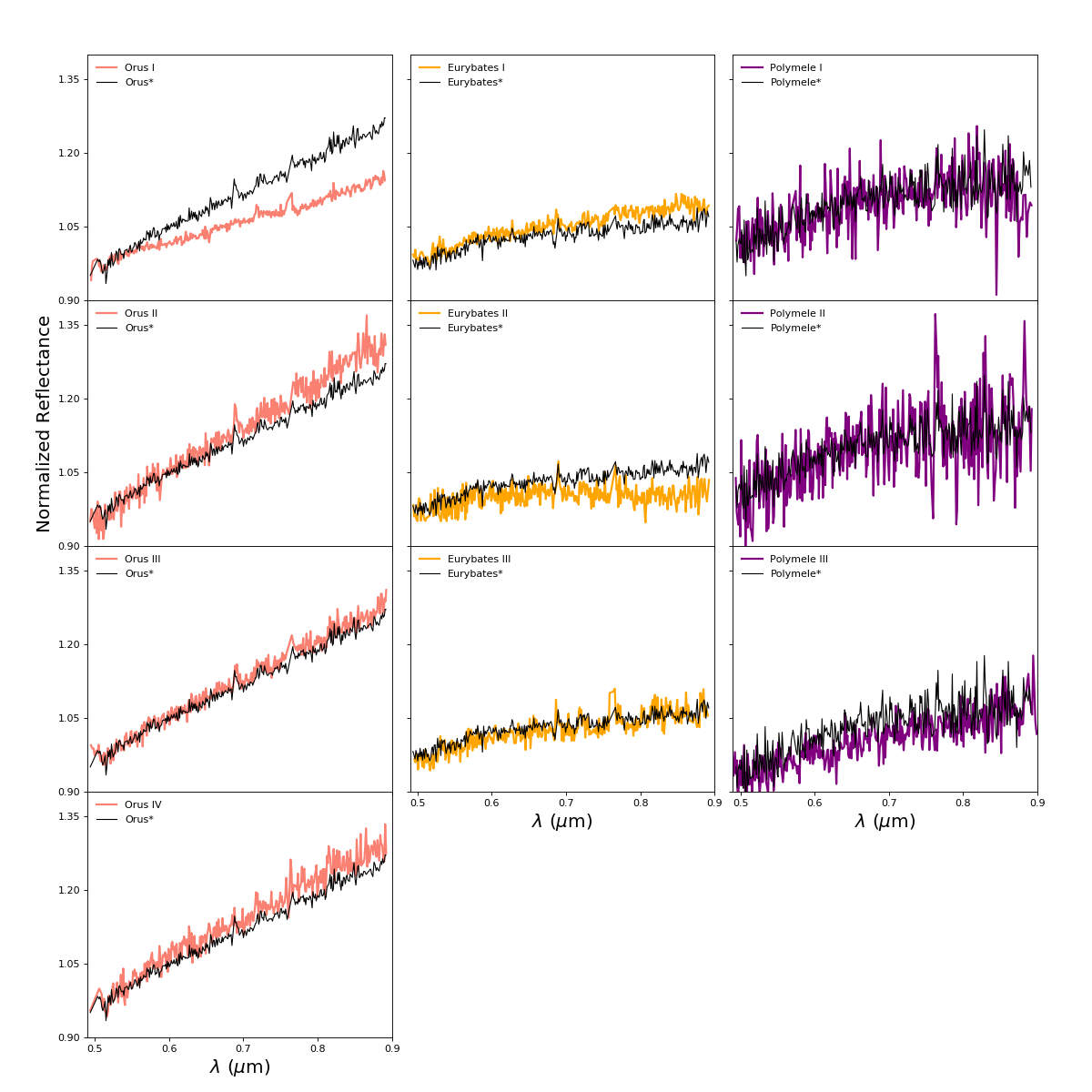} 
	\end{center}
	\caption{Reflectance spectra of our sample normalized to unity at 0.55 $\mu$m. Spectra of Orus (salmon), Eurybates (orange), and Polymele (purple). The black spectrum in each box represents the combined spectrum of each target normalized to unity at 0.55 $\mu$m.}
	\label{fig1}
\end{figure*}

\section{Observation and Data Reduction}\label{dois}

\subsection{Spectroscopic observations}

Details of spectroscopic observations conditions for each object are listed in Table \ref{tab1} and of solar analogs stars in Table \ref{tab2}. Information includes objects name, magnitude, date of observation, starting UT, exposition time, and airmass. The spectra of the MBA were obtained on May 26 at the 10.4 m GTC at the El Roque de Los Muchachos Observatory (ORM) in La Palma, Spain. JTs spectra were taken on July 7 and August 12 at the 4.1 m SOAR Telescope on Cerro-Pach\'on, Chile.   

We used the Optical System for Imaging and Low Resolution Integrated Spectroscopy (OSIRIS) camera spectrograph \citep{cepa2000, cepa2010} of GTC to observe the MBA. We used the R300R grisma with a slit of 1.23" and the $2 \times 2$ binning mode with a readout speed of 200 kHz, which corresponds to the standard operation mode of the instrument. In all observation nights, the target were oriented to the parallactic angle to minimize losses due to atmospheric dispersion. We took 3 consecutive spectra of Donaldjohanson in a range of 0.49 to 0.92 $\mu$m and shifted them into the slit direction by 10" to improve the sky subtraction and fringing correction. To observe the JTs, we used the Red Camera of Goodman High Throughput Spectrograph (GTHS) of 4.1m SOAR with the grating of 400 lines/mm, a slit of 1.50" with GG455 second order blocking filter, and a readout speed of 344 kHz, which provides an effective spectral interval from 5000 to 9050 {\AA} with a dispersion of 1.0 {\AA}/pix. We used this same configuration to obtain all JTs spectra. In total, we did 3 observations of Eurybates and Polymele, and 4 of Orus, in which each observation contains 3 consecutive spectra of each target (with exception to the second observation of Polymele that contain only 2 consecutives spectra).

We applied standard reduction techniques using IRAF\footnote{IRAF is distributed by the National Optical Astronomy Observatories, which are operated by the Association of Universities for Research in Astronomy, Inc., under cooperative agreement with the National Science Foundation.} routines for all targets. Images were initially bias and flat-field corrected, the sky background was subtracted to them. We extract the spectra using an aperture that varied depending on the seeing of the observation. The spectra were wavelength calibrated with HgArNe (SOAR data) and XeNeHgAr (GTC data) lamps. To correct for telluric absorption and obtain the relative reflectance, we observed solar analogs stars from the Landolt list \citep{landolt1992}, using the same instrument configuration, and at similar airmasses as to the objects. We observed 2 solar analog stars on May 26, 6 on July 7, and 2 on August 12. 

%\textcolor{green}{Different solar analogs produced differences in the reflectance spectra.} These differences could propagate into the spectrum of the target through the reduction process. To quantify the errors, we divided the solar analogs stars spectra by one that we take as reference (the one with intermediate airmass). The result of division should be a straight line with spectral slope $S^\prime = 0$, but the analysis shows a variation of 0.50$\%/1000$ {\AA} for the night of May 26, 0.84$\%/1000$ {\AA} to the night of July 7 and 1.1$\%/1000$ {\AA} to the night of August 12. Each individual spectrum of the objects was then divided by the corresponding mean spectra of the solar analogs obtained in each observation night, then averaged and normalized to unity at 0.55 $\mu$m. 

%In order to improve the signal-to-noise ratio (SNR) of the spectra in our sample, we applied a binning to each spectrum, with a binsize of 11 points. Then\textcolor{red}{,} we \textcolor{red}{replaced} the reflectance value corresponding to the central wavelength of the \textcolor{red}{bin} by the median reflectance value to avoid outliers and to make the resulting spectrum more robust. We estimate the SNR of each spectrum to have a parameter to evaluate quality of the spectra.
Finally, each object spectrum was divided separately by the spectrum of the solar analog stars, normalized to unity at 0.55 $\mu$m and binned with a bin size of 11 points. These final relative phase reflectances of the JTs are shown in Fig. \ref{fig1} and the spectrum of the MBA in Fig. \ref{classification}. We also show in Fig. \ref{fig1} the relative reflectance of each target obtained combining all the individual spectra during the whole observing period (we will refer to this as the combined spectrum).

\subsection{Photometric observations}

Photometric observations of Eurybates were obtained at the OASI, Brazil. We used the 1.0-m Telescope with a 2k$\times$2k CCD, which provides a field of view of 11.8' and an image scale of 0.343"/pixel. Observations were performed on October 3 and 4 of 2018. The average seeing of both nights was approximately 3.5". The LCs were obtained in the \textit{r} and \textit{i} filters of the Sloan Digital Sky Survey (SDSS) system. We applied standard reduction techniques in both filters: images were bias, dark and flat-field corrected. To correct for fringing patterns, present in images taken with \textit{i} filter, we performed a dithering sequence between expositions. These images were combined to build a mask that was later used to correct the individual exposures. 

Relative photometry was computed to obtain the rotational LC in both filters. We tested different apertures in order to maximize the object's SNR in each night. We also tried different radius and widths for the annulus used for the background subtraction. We chose the same reference stars set during the two observation nights. The LCs in \textit{r} and \textit{i} filters are presented in Fig. \ref{foto}.

\section{Analysis and Results} \label{tres}

We obtained four spectra of Orus, and three of Polymele and Eurybates that cover 48\%, 85\% and 63\% of a whole rotation, respectively. All of these spectra are reddish and featureless, as expected for primitive surfaces (see Fig. \ref{fig1}). We obtained one spectrum of Donaldjohanson (see Fig. \ref{classification}). To search for variations in the spectral characteristics with the rotation, we will use the spectral slope and the taxonomical classification. Some variations in color are already suggested in the figure, however, it is well known that small differences can be caused by many different factors such as different compositions or particle sizes on the asteroid, space weathering, or even more common, observational uncertainties associated with the observing conditions. Because of this, we also show and characterize the relative reflectance of each target obtained merging all the individual spectra during the whole observing period (we will refer to this as the combined spectrum). Below we show the results of the analysis of these spectra that are also included in Tab. \ref{fim}, with the physical properties of them.

\subsection{Jupiter Trojans targets: Eurybates, Orus and Polymele}

To search for rotational variability in the JTs spectra, we selected an arbitrary zero-point for the rotational phase as the first spectrum of each object. The remaining acquisitions were rotationally calibrated in comparison with the first. The physical properties of each object can be seen in Tab. \ref{fim}. We measured the visible spectral slope ($S^\prime $) for each spectra in our sample and looked for variations higher than 1-$\sigma$ level.

\begin{figure*}[!t]
	\centering
	\includegraphics[scale=0.55]{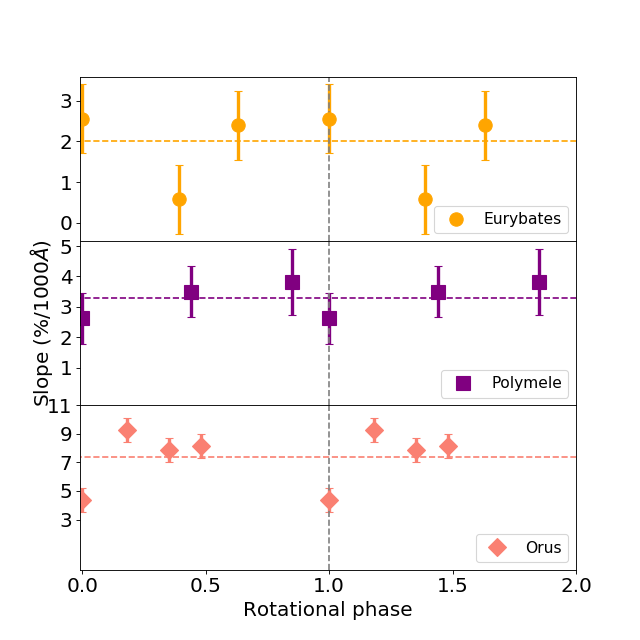}
	\caption{Slopes of the observed objects versus rotational phase. Each point represents the slope of an object with the errorbar of 1-$\sigma$ level and the dashed line in each panel represents the slope of the combined spectrum of each target. Eurybates is shown in orange circles, Polymele in purple squares and Orus in salmon diamonds. The spectral slopes are in units of \%$/1000${\AA} and the values of the rotational phase are arbitrary with 0 representing the beginning of the spectroscopic observations and 1, the end of the rotational period of the object. The rotational period is duplicated only to help the visualization.}
	\label{slop}
\end{figure*}

We used the definition of $S^\prime $ presented in \cite{jewitt90} to calculate the spectral slope. We performed a linear fit using the wavelength range between 0.5 and 0.9 $\mu$m, where the reflectance is well represented by a line fit, and normalized it in 0.55 $\mu$m. The values for the slope and the associated uncertainty are presented in units of $\%/1000${\AA}. The values obtained for the slope of each spectrum and for the combined spectrum are shown in Tab. \ref{tab4}, along with the associated error, and plotted in Fig. \ref{slop} with 1-$\sigma$ errorbars for each spectrum. 

The errors in the slope values are computed following the law of propagation of uncertainty, taking into account the error in the fit and the error introduced by dividing by the spectra of the solar analog stars or the centering in the slit. These differences could propagate into the spectrum of the target through the reduction process. To quantify the errors, we divided the solar analogs stars spectra by one that we take as the reference (the one with intermediate airmass). The result of division should be a straight line with spectral slope $S^\prime = 0$, but the analysis shows a variation of 0.50$\%/1000$ {\AA} for the night of May 26, 0.84$\%/1000$ {\AA} of the night of July 7 and 1.1$\%/1000$ {\AA} of the night of August 12. We took these values as systematic errors. We also evaluated the error in the fit for each phase to calculate the slope, and this error was added quadratically to the systematic error. The final error is strongly dominated by the systematic one.

\begin{table*}[t!]
\centering
\begin{tabular}{ccccccc}
\hline 
\hline
Object & \textit{a} (AU) & \textit{e} & \textit{i}($^\circ$) & $p_\nu$ & Rot. Period (hr) & Diameter (km)\\
\hline
Eurybates & 5.194 & 0.088 & 8.059 & $0.052 \pm 0.007$ + & $8.702724 \pm 0.000009$ * & $63.8 \pm 0.3$\\
Polymele & 5.164 & 0.0947 & 12.989 & $0.09 \pm 0.02$ + & $5.87607 \pm 0.0005$ - & $21.0 \pm 0.1$\\
Orus & 5.125 & 0.0362 & 8.468 & $0.07 \pm 0.01$ + & $13.48617 \pm 0.00007$ * & $50.8 \pm 0.8$\\
Donaldjohanson & 2.383 & 0.186 & 4.423 & $0.10 \pm 0.02$ \# & - & $3.895$\\ 
\hline
\end{tabular}
\caption{Physical properties of the sample: semi-major axis (\textit{a}), eccentricity (\textit{e}), inclination (\textit{i}), visible albedo ($p_\nu$), and rotation period (Rot. Period). The values of \textit{a}, \textit{e}, and \textit{i} were taken from JPL Small-Body Database Browser. References: (+) \cite{grav2012}, (*) \cite{moto}, (-) \cite{buie2018light}, and (\#) \cite{masiero2011} }
\label{fim}
\end{table*}

In addition to the slope, we also performed a taxonomical classification in each single spectrum and also in the combined spectrum. We used the CANA\footnote{The CANA toolkit (Codes for ANalysis of Asteroids) is a Python package specifically developed to facilitate the study of features in asteroids spectroscopic and spectrophotometric data.} package for this task \citep{cana}. The routine applied by CANA is based on minimizing the chi-squared between the asteroid spectrum and the taxonomy templates \citep{demeo09}. Although the aforementioned taxonomic system is an extension to the NIR of the \cite{busebinzel} taxonomy in the visible, all the obtained classes coexist in both schema. The classification results are shown in Tab. \ref{tab4}.

\begin{table*}[h!]
	\small
	\centering
	\begin{tabular}{ccccc}		
		\hline
		\hline
		Object & Rot. Phase &  $S^\prime$($\%/1000${\AA})  & Tax. & SNR\\
		\hline
		Eurybates-I & 0.00& 2.56 $\pm$ 0.84 & Xc & 62\\
		Eurybates-II & 0.39 & 0.58 $\pm$ 0.84 & C & 50\\
		Eurybates-III & 0.63& 2.39 $\pm$ 0.84 & Xc & 44\\
		Combined & & 2.00 $\pm$ 0.84 &  X$_k$ & 91\\ 
		\hline
		Polymele-I &  0.00 & 2.63 $\pm$ 0.84 &  X$_k$ &23\\
		Polymele-II & 0.44 & 3.49 $\pm$ 0.84 &  X$_k$ &18\\
		Polymele-III & 0.85 & 3.81 $\pm$ 1.1 &  X$_k$ &31\\
		Combined &  & 3.29 $\pm$ 0.92 & X$_k$ & 42\\
		\hline 
		Orus-I & 0.00 & 4.40 $\pm$ 0.84 & Xk &48\\
		Orus-II & 0.18 & 9.27 $\pm$ 0.84 & D & 42\\
		Orus-III & 0.35 & 7.84 $\pm$ 0.84 & D & 51\\
		Orus-IV & 0.48 & 8.13 $\pm$ 0.84 & D & 33\\ 
		Combined & & 7.36 $\pm$ 0.84 & D & 60\\ 
		\hline
		Donaldjohanson &  & 0.36 $\pm$ 0.50 & Cg & 57\\
		\hline
	\end{tabular}
	\caption{Rotational phase (Rot. Phase), visible spectral slope ($S^\prime$), taxonomy classification (Tax.), and estimated signal-noise ratio (SNR) of each spectrum of the sample and the combined spectrum of each object.}
	\label{tab4}
\end{table*}

The spectra of all the JTs in our sample are featureless. We covered 85\% of the rotation period of Polymele with 3 spectra and the spectral slope variation does not exceed the 1-$\sigma$ level compared with the slope of the combined spectrum of 3.29 $\%/1000${\AA}. The 4 spectra of Orus covered 48\% of its rotation period. The first spectrum of Orus shows a clear outlier spectral slope of 4.4 $\pm$ 0.84 $\%/1000${\AA}, less red than the other three with values between 7.84 and 9.27 $\%/1000${\AA} (see Fig.\ref{fig1} and Tab.\ref{tab4}). These three spectra and the combined are classified as D-types, meanwhile the first one is classified as X$_k$. In this specific case, we attribute the slope variation to instrumental setup and climatic conditions in the moment of the first observation. In Sect. \ref{cinco}, we discuss this in more detail. The analysis of the three spectra of Eurybates revealed that the slope of the second spectrum, classified as a C-type, deviates from the first and the third that were classified as Xc-types (see Tab. \ref{tab4}). The slope deviation is above 1-$\sigma$ level. We closely inspected the observational circumstances, and we find that the spectral variability must correspond to real variations on the surface of Eurybates. 

\begin{figure*}[t!]
	\begin{center}
		\includegraphics[scale=0.4]{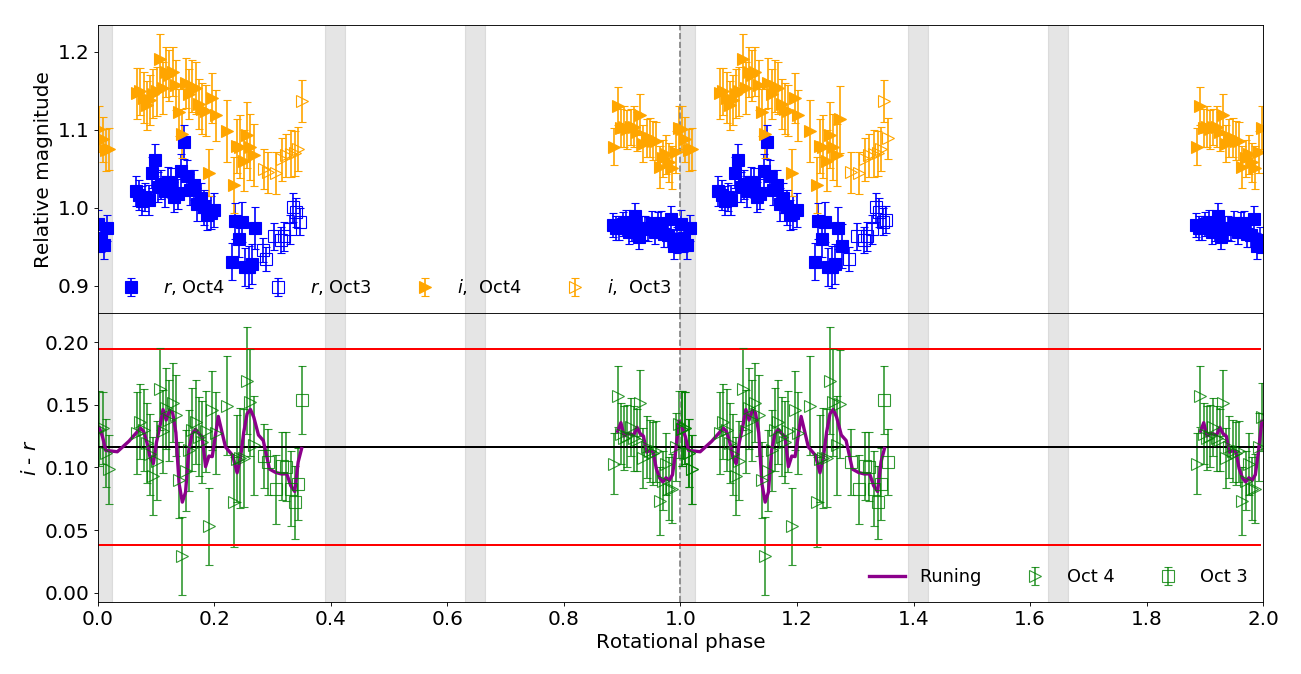}
	\end{center}
	\caption{\textit{Top panel:} Photometry in the \textit{r} (blue) and \textit{i} (orange) filters of Eurybates obtained at OASI. The filled symbols represent the objects observed in October, 4 and the empty ones in October, 3. \textit{Bottom panel:} Difference between the \textit{i} and \textit{r} filters (green). The black horizontal line shows the mean difference and the red ones, the 3-$\sigma$ standart deviation. A runing median (width = 3) is overplotted as a thick purple line. The horizontal axis represents the rotational phase. We choose the 0 point at the same time of the spectroscopic observations to be able to compare both results. The gray regions are the period covered by the spectroscopic data. The rotational period is duplicated only to help the visualization.}
	\label{foto}
\end{figure*} 

To confirm this, we conducted a photometric analysis similar to the ones presented in \cite{lacerda2008} and \cite{lacerda2009}. The LCs in \textit{r} and \textit{i} filters, are presented separately in Fig. \ref{foto}-\textit{top}. Since simultaneous measurements in both filters were not possible, we cross-interpolated their fluxes of the \textit{r}-band taking the \textit{i}-band time coordinates as reference. We subtracted the respective magnitudes in each filter to produce the objects' color curve (Fig. \ref{foto}-\textit{bottom}). The resulting color curve was smoothed using a running median to search for structures that could be associated with a rotational color variability. However, we found no variation that exceeded the 2-$\sigma$ level. Note that our photometric data covered $\sim35\%$ of Eurybates' rotation period, and did not reach the same phase observed by the spectroscopic observations. More data is warranted to secure confirmation. 

\subsection{Donaldjohanson}

\begin{figure}[h!]
	\centering
	\includegraphics[scale=0.38]{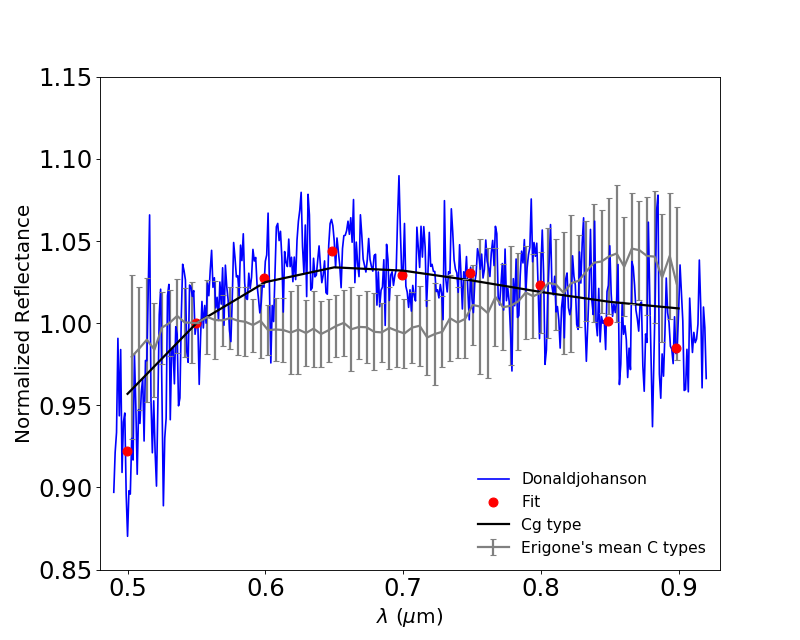}
	\caption{Comparison of the normalized spectrum (at 0.55$\mu$m) of Donaldjohanson (blue) with the Cg-type (black) from Bus-DeMeo taxonomy and the mean of C-types of (163) Erigone family (grey). The red dots are the fit of Donaldjohandon spectrum.}
	\label{classification}
\end{figure}

In the case of Donaldjohanson, which only has one spectrum we did not search for rotational variability. The spectrum of Donaldjohanson presented a decay of reflectance towards the ultra-violet wavelengths (Fig. \ref{classification}). This feature is commonly observed in C-type asteroids. \cite{vilas1995ub} suggested that it could be related to hydrated materials in the surface of the asteroid. We characterized the turnover region following the methodology described in \cite{de2018primass}; the turn-off point was measured by calculating the furthest point from a linear fit considering just the edges of the spectrum. In addition to the turn-off wavelength, we also measured the slope of the drop-off in reflectance on wavelengths lower than the turn-off point (near-UV slope).

We found a turn-off point at 0.6206 $\pm$ 0.0038 $\mu$m, with a near-UV slope of 9.32 $\pm$ 0.50 $\%/1000${\AA}. The visible slope $S^\prime$ is 0.36 $\pm$ 0.50 $\%/1000${\AA}, and is classified as a Cg-type. In Fig. \ref{classification}, we show the spectrum of the object in comparison with the Cg-class template and the mean spectrum of the C-types within the Erigone family from \cite{morate16}.

\section{Discussion}\label{cinco}

\subsection{The spectral properties of our sample}

We observed 3 JTs and one MBA that are targets of NASA Lucy Mission. The sample presented in this work shows spectra compatible with primitive asteroids, with colors ranging from gray to red. We did not find any strong evidence of variability on the spectra of Polymele. Spectral slope variations can arise from systematic observational effects, often related to seeing conditions \citep{binzel2015}. In the case of the first spectrum of Orus, which was acquired in the beginning of the night, the seeing was higher than the slit adopted. Therefore, we consider that the variability in the associated slope could be the result of a systematic error. All of the three remaining spectra, which were observed with a considerably smaller seeing, presented a self-consistent behavior. 

\begin{figure}[h!]
\centering
\includegraphics[scale=0.33]{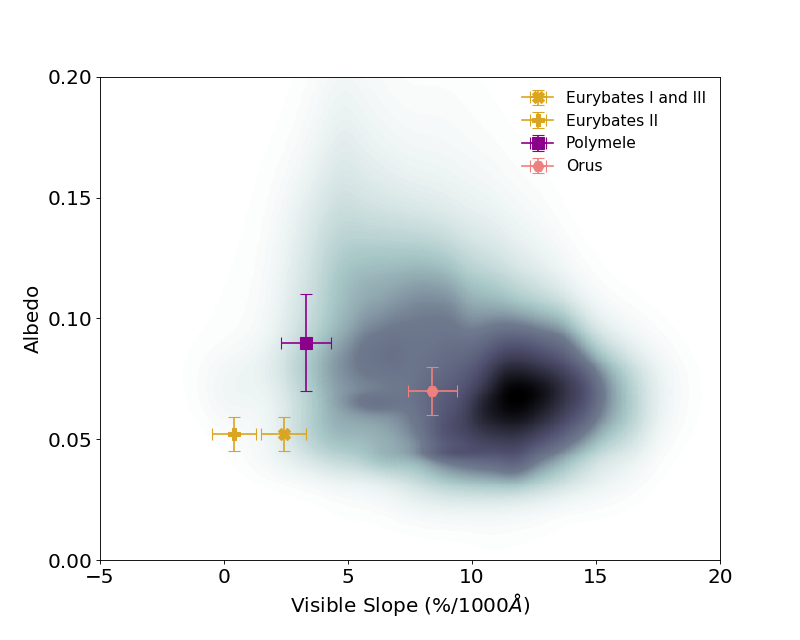}
\caption{Density plot of JTs: albedo \textit{versus} visible spectral slope ($\%/1000${\AA}) and highlighted, the JTs in our sample. The errorbars represent the sistematic error in 1-$\sigma$ level.}
\label{comparison}
\end{figure}

Figure \ref{comparison}, adapted from \cite{de2018primass}, shows a compilation of NEOWISE and SDSS properties as visible slope and geometric albedo for JTs. The spectra of Orus and Polymele present rather homogeneous characteristics along the surfaces, and their taxa correspond with those of the two dominant populations in the Trojan population, the P- (low-albedo X type asteroids) and the D-type group of objects (see Fig. \ref{comparison}). These taxonomic classes are observed across the main belt, but they become predominant with increasing heliocentric distance. However, when looked closely in the Albedo \textit{versus} Spectral slope (Fig. \ref{comparison}) space of parameters, these two Lucy targets reveal as peculiar. Polymele’s color is slightly less red than the average color of the P-type Trojans, meanwhile, Orus, classified as a D-type, is considerable more neutral than the average D-type Trojans, as well as the typical D-type Hildas and Cybeles, \citep{de2018primass}. Moreover, its albedo is higher than the albedo of these D-type suggesting that the surface properties of this trojan may be comparable with those of the members of the P-type trojans. This knowledge will shed light on how the JTs were formed and on the history of the Solar System. 

According to dynamical evolution models, the resonant populations in the outer main belt and up to the Neptune orbit, namely the Hildas, JTs, Centaurs and Neptunian Trojans should have similar origin \citep{morbidelli2005chaotic,broz2008} but observations show that their surface characteristics are differents \citep{jewitt2018trojan, de2018primass}. The fact that they show different properties suggest that either there is a thermophysical effect operating at their regions, or they have distinct origin. The Lucy mission will investigate the surface of Orus and Polymele, and will reveal if they have distinct compositions, and possibly different origins. 

The other two objects in our study correspond to the C-complex, a class that dominates the inner belt but that is unusual in the Trojan population. It was expected that Donaldjohanson was a C-type asteroid for belonging to the Erigone collisional family \citep{nesvorny2015}, a primitive family located in the inner part of the main belt that has been characterized within the PRIMitive Asteroids Spectroscopic Survey (PRIMASS, see \cite{morate16}). Nearly 50$\%$ of the family was classified as C-type asteroids, with more than 80\% of these objects presenting an absorption band in their visible spectra, centered at 0.7 $\mu$m \citep{morate16}. This feature is associated with silicates that have been altered by the presence of liquid water. The spectrum of Donaldjohanson does not show such a feature. Accordingly with our analysis, it is classified as a Cg-type, a sub-type of C-complex with a pronounced UV drop-off.

Similar to Donaldjohanson, Eurybates has one spectrum that was also classified as a C-type and does not show hydration band. Interestingly, the C-type spectrum of Eurybates does not show the UV drop of reflectance. Note that such a feature has been identified in four out of 17 members of the Eurybates family \citep{fornasier2007visible}. This feature is detected on the spectra of many main belt C-type asteroids \citep{vilas1994, vilas1995ub}, and it is often associated with other spectral features due to aqueous alteration products. While it is likely that Donaldjohanson possess hydrated minerals on its surface, due to the association with the Erigone family, it is not clear if Eurybates presents such materials. The observations that will be performed by the Lucy mission will provide valuable knowledge on the composition of both objects and how the surface of a C-complex asteroid can vary.% It will also show whether or not the UV-drop of reflectance is in fact related to hydrated minerals.

\subsection{Does Eurybates have an inhomogeneous surface?}

Although the attempt to detect rotational color variation in Eurybates using photometric techniques was inconclusive, due to the short fraction of the period that was covered, we found spectroscopic slope variation that possibly indicate variability on its surface. In addition to the described analysis of the Eurybates spectra acquired in this work, we also analyzed the spectrum presented in \cite{fornasier2007visible}, which was obtained at a different phase angle. They measured a slope of (-0.18 $\pm$ 0.57\%$/1000${\AA}). However, they used a different spectral range to calculate the spectral slope. For this reason, we recalculated it following the methodology described in section \ref{tres}. The variability in the slope presented in Tab. \ref{tabn} is above the 1-$\sigma$ deviation, considering the uncertainties. 

\begin{table}[!h]
	\centering
	\begin{tabular}{ccc}
		\hline
		Eurybates &  $S^\prime$(\%$/1000${\AA}) & $\alpha$($^\circ$)\\
		\hline
		F07 & 0.15 $\pm$ 0.50 & 0.7848\\
		This work I & 2.56 $\pm$ 0.84 & 5.3382\\
		This work II & 0.58 $\pm$ 0.84 & 5.3144\\
		This work III & 2.39 $\pm$ 0.84 & 5.3007 \\
		\hline
	\end{tabular}
	\caption{Spectral slope ($S^\prime$) and phase angles ($\alpha$) of Eurybates found in literature from \cite{fornasier2007visible} recalculated as outlined in the text, in comparison with our results.}
	\label{tabn}
\end{table}

The different slope values for Eurybates obtained in this work are probably not caused by phase reddening, an effect which has been deeply studied in the literature \citep{lumme81, luu1990, clark2002, nathues2010, sanchez2012, reddy2012}. Although the spectrum observed in \cite{fornasier2007visible} had a lower phase angle, the associated slope is in agreement with the second Eurybates spectrum obtained in this work, which was acquired at similar phase angle as the first and third spectrum (Tab. \ref{tabn}). Even if the particular case of the second spectrum was discarded, the variation would still be too high to be explained only by phase reddening effect. \cite{lumme81} measured a change in spectral slope of 0.15\%$/1000${\AA} per degree for a sample of C-type asteroids. In our case, we found a variation in Eurybates' spectral slope between the mean of our redder spectra and that from \cite{fornasier2007visible} of 2.32\%$/1000${\AA} for a change in phase angle of 4.5$^\circ$, i.e., a variation of 0.51\%$/1000${\AA} per degree, which is considerably larger than the aforementioned work. 

A more likely possibility is that the surface of Eurybates is not entirely homogeneous. Eurybates is the largest body of the homonyms family, and has undergone a strong collision in the past. \cite{fornasier2007visible} conducted an analysis with 17 members of the family and found that their slope vary from -0.5 to 4.6 \%$/1000${\AA}, which makes the family predominantly composed by C- and P- type asteroids. The longitudinal variability in the slope of Eurybates is smilar to the dispersion within the family. This result, combined with the fact that these low values of slope are not commonly observed outside of the family \citep{fornasier2007visible, wong2014, de2018primass}, could suggest that physical properties of Eurybates are related to its collisional history.% \cite{dotto2008troianis} proposed that this family could have been originated by the fragmentation of a peculiar body or, alternatively, could be an old family, where space weathering processes have covered any original differences in composition among the different members. However, besides the age of the family is not well determined, it is not clear how the break-up of a primitive body, that could potentially hold pristine volatile material in the inner layers, would affect the physical properties of the family members surface.

The longitudinal variation of Eurybates slope could be caused by different factors: regions with different ages, regolith size, composition, or a combination of these factors. A collision could alter the properties of only a localized region of the body, and also excavate and expose inner-layer material. Areas of pristine material excavated by impacts might have much higher albedo than the $\sim$ 5\% typical of Trojans \citep{fernandez2003}. Likewise, these newly-exposed regions might have a distinct color from the rest of the radiation-reddened surface \citep{wong2016}. Other factor that could affect the spectral slope is distinct regolith sizes on the surface of Eurybates. Particle size affects the slope of overall reflectance. This behavior was proposed as one of the possibilities to explain spectral variation of 101955 Bennu \citep{binzel2015}.

\section{Conclusions}\label{seis}

We obtained visible spectra of the targets of the Lucy mission, including 3 JTs with the Goodman SOAR spectrograph and one MBA with the OSIRIS camera-spectrograph at GTC. In this work, we show, for the first time, a rotational spectral analysis in the visible for these JTs. For Donaldjohanson, we also performed for the first time the spectral characterization in the visible. In general, we did not find any absorption band on the analyzed spectra of our sample.

\begin{itemize}

\item Eurybates: We covered 63\% of the rotation period of this object and we compared the spectral slopes obtained in our work with those already published in the literature. This comparison reinforced our hypothesis that the observed variations in the slope may be related to heterogeneities on the surface of this object. Such variation could be explained by the existence of an associated collisional family, originated during a cratering-forming event, that might have exposed fresh material, or by the presence of distinct regolith sizes on the surface of the object.  

\item Orus and Polymele: The spectral slopes and taxonomic classification of these objects do not present hints of surface variability on their surfaces. While representatives of the D- and P- taxonomic types, respectively, our data reproduced the spectral characteristics of the taxonomic groups. 
%\item Orus: We covered 48\% of the rotational period of Orus. Due to the observational circumstances in which the first spectrum was obtained, we are not in a position to assert if the variation noted between its first spectral slope and the others is real. We need to observe it again for a more reliable interpretation of our results.
\item Donaldjohanson: It is the only MBA that will be visited by the Lucy mission. Unlike the spectra of the JTs, the spectrum of this object shows a drop-off in reflectance in the region below 0.62 $\mu$m. Although almost 90\% of the C-types asteroids of the family to which it belongs have a band centered at 0.7 $\mu$m, related with hydrated materials, it does not possess it. All these characteristics place it in the taxonomic group of Cg types. Lucy will provide a big picture of C-class asteroids: members of dynamical families with distinct size in different regions of the Solar System. 
\end{itemize}

Overall, spectroscopy of the Lucy targets covered in this work show diversity in the surface properties of these primitive asteroids. No clear surface variations with rotation are detected for Polymele and Orus. The combined spectrum and albedo of these bodies show that their properties are in some way peculiar when compared with the dominant characteristics of the P- and D- dominant groups in the whole Trojan population. Spectroscopy of Eurybates, on the other side, suggests that some variation on the characteristics of the reflectance of this body could be related with its collisional history. If this is related to its composition or to the particle size distribution is something that will be further explored by the Lucy mission, that can now plan the observational strategy to cover the different regions on the surface of Eurybates, according to our results. Finally, Donaldjohanson, the only main belt object in the group of targets, shows, according to our visible spectrum, hints of the presence of hydrated materials. This fact, could be explored with additional data in the 3-microns region from ground-based facilities or using the James Webb Space telescope instrumentation, in particular NIRSpec will be ideal to search not only for the -OH absorption band but also for the complex absorption due to organic components expected to dominate the surface of primitive objects in the Solar System \citep{2016riv, 2019noe}.

\section*{Acknowledgments}

The authors are grateful to the IMPACTON team, in particular T. Rodrigues, D. Lazzaro, R. Souza and A. Santiago, the first two for keeping the OASI operative and the others for the technical support. The whole team of the SOAR and GTC Telescopes. This study was financed in part by the Coordena\c{c}\~ao de Aperfei\c{c}oamento de Pessoal de N\'ivel Superior - Brasil (CAPES) - Finance Code 001. Souza-Feliciano would like to thank Josh Emery for kindely providing some of the data used in this work. AAC acknowledges support from FAPERJ (grant E26/203.186/2016) and CNPq (grants 304971/2016-2 and 401669/2016-5). M. De Pr\'a and E. Fern\'andez-Valenzuela acknowledge funding through the Preeminant Postdoctoral Program of the University of Central Florida. NPA acknowledges support from SRI/FSI funds through the project "Earth- and Space-Based Studies in Support of NASA Space Missions". J. de Le\'on acknowledges financial support from the projects SEV-2015-0548 and AYA2017-89090-P (MNECO).   
%\textcolor{red}{Partially based on observations obtained at the Southern Astrophysical Research (SOAR) telescope, which is a joint project of the Minist\'erio da Ci\^encia, Tecnologia, Inova\c{c}\~ao e Comunica\c{c}\~oes (MCTIC) da Rep\'ublica Federativa do Brasil, the U.S. National Optical Astronomy Observatory (NOAO), the University of North Carolina at Chapel Hill (UNC), and Michigan State University (MSU). Another part of this work is based in observations acquired at GTC (ORM-La Palma) and, in observations made at Observat\'orio Astron\^omico do Sert\~ao de Itacuruba (OASI).}

\section*{References}
\bibliographystyle{elsarticle-harv} 
\bibliography{mybibfile}

\end{document}